\pdfoutput=1

\documentclass[nofootinbib,twocolumn,superscriptaddress,aps,preprintnumbers,amsmath,amssymb,prd]{revtex4-1}

\usepackage{varioref}
\usepackage{hyperref}
\usepackage{cleveref}

\usepackage{epsfig}
\usepackage{amsmath}
\usepackage{amssymb}
\usepackage{subfigure}
\usepackage{cancel}
\usepackage{textcomp}
\usepackage{calc}
\usepackage{graphicx}
\usepackage{dcolumn}
\usepackage{color}
\usepackage{xcolor}
\usepackage{pifont}
\usepackage{slashed}
\usepackage{fdsymbol}

\usepackage{footnote}
\usepackage{hyperref}
\definecolor{linkcolor}{rgb}{0.0, 0.6, 0.7}
\definecolor{Green}{rgb}{0.0, 0.7, 0.4}
\hypersetup{
  colorlinks=true,
  urlcolor   = linkcolor,
  linkcolor  = linkcolor,
  citecolor  = linkcolor
}
\newcommand{\be}{\begin{equation}}
\newcommand{\ee}{\end{equation}}
\newcommand{\eea}{\end{eqnarray}}
\newcommand{\bea}{\begin{eqnarray}}

\definecolor{JMLB}{RGB}{4, 55, 242}

\begin{document}

%%%%%%%%%%%%%%%%%%%%%%%%%%%%%%%%%%%%%%%%%%%%%%%%%%%%%%%%%%%%%%%%%%%%%%%%%%%%%%%%%%%%%%%%%%%%%%%%%%%%

\preprint{CERN-TH-2023-117}
\preprint{DESY 23-095}

\title{Implications of Protecting the QCD Axion in the Dual Description}
%Clockworking Natural Inflation

\author{Gongjun Choi}
\email{gongjun.choi@cern.ch}
\affiliation{Theoretical Physics Department, CERN, 1211 Geneva 23, Switzerland}

\author{Jacob Leedom}
\email{jacob.michael.leedom@desy.de}
\affiliation{Deutsches Elektronen-Synchrotron DESY, Notkestr. 85, 22607 Hamburg, Germany}

%%%%%%%%%%%%%%%%%%%%%%%%%%%%%%%%%%%%%%%%%%%%%%%%%%%%%%%%%%%%%%%%%%%%%%%%%%%%%%%%%%%%%%%%%%%%%%%%%%%%

\begin{abstract}
The QCD axion can be be formulated in a dual description as a massive 2-form field. In this picture, the QCD axion quality problem translates into the question if there are additional 3-forms coupled to the axion other than the QCD 3-form that emerges at low energy. If such forms exist, the quality problem can be resolved via the introduction of other massive 2-forms (and thus corresponding axions), one for each additional 3-form. This can motivate an “axiverse from a high quality QCD axion". In this work, we discuss this issue in the general case where the QCD axion couples to arbitrarily many 3-forms. Given the multiple axion solution, we discuss the phenomenological implications of the enhanced quality of the QCD axion in the dual description. These include sub-eV axion-like particle search through the axion-photon coupling, the cosmological consistency of a large decay constant QCD axion, and a model for the observed cosmic birefringence.
\end{abstract}

%%%%%%%%%%%%%%%%%%%%%%%%%%%%%%%%%%%%%%%%%%%%%%%%%%%%%%%%%%%%%%%%%%%%%%%%%%%%%%%%%%%%%%%%%%%%%%%%%%%%

\date{\today}
\maketitle

%%%%%%%%%%%%%%%%%%%%%%%%%%%%%%%%%%%%%%%%%%%%%%%%%%%%%%%%%%%%%%%%%%%%%%%%%%%%%%%%%%%%%%%%%%%%%%%%%%%%
\section{Introduction}
The QCD axion provides an elegant resolution to the Strong CP problem present in the Standard Model (SM)~\cite{Peccei:1977hh,Peccei:1977ur,Weinberg:1977ma,Wilczek:1977pj}. In brief, the Standard Model supplemented with the QCD axion contains the operators
\bea
    \mathcal{L} \supset \frac{1}{32\pi^2}\bigg(\theta + \frac{a}{F_a}\bigg) \text{Tr}[F_{\mu\nu} \widetilde{F}^{\mu\nu}]
\eea
where $\theta$ encodes contributions from a bare parameter as well as phases of the colored fermion mass matrix,  $F^a_{\mu\nu}$ is the $SU(3)_c$ field strength tensor with $\widetilde{F}^{a\mu\nu}=(1/2)\epsilon_{\mu\nu\rho\sigma}F^{a\rho\sigma}$ its dual, and $a$ is the QCD axion with $F_a$ its decay constant. Measurements of the neutron electric dipole moment find a constraint on the effective theta parameter $\bar{\theta}\equiv\theta+(a/F_{a})$ as $\bar{\theta}\lesssim10^{-10}$~\cite{Baker:2006ts}. In the absence of an axion, this result indicates an extreme fine tuning in the SM. However, the axion provides a natural explanation -- due to the above Chern-Simons coupling, the axion develops a potential from QCD instanton effects and $\bar{\theta}$ is dynamically set to zero. 

In spite of this appealing picture, the QCD axion is plagued by a \textit{quality problem}: if there are additional contributions to the axion potential beyond QCD instanton effects, then the minimum $\bar{\theta}_{\rm min}$ of the total potential for $\bar{\theta}$ may violate $\bar{\theta}\lesssim10^{-10}$. Thus the QCD axion would no longer resolve the Strong CP puzzle. 

The global nature of the $U(1)_{\rm PQ}$ Peccei-Quinn (PQ) symmetry~\cite{Peccei:1977hh,Peccei:1977ur} is one of underlying reasons why the QCD axion is at risk. Since global symmetries are generically expected to be broken in an ultraviolet unification of the SM and gravity~\cite{Hawking:1987mz,Giddings:1988cx,Banks:2010zn,Harlow:2018tng,Harlow:2018jwu}, PQ symmetry violating higher dimensional operators such as $\mathcal{L} \supset c_{n}(\Phi^{n}+\Phi^{\dagger\, n})/M_{P}^{n-4}$ ($n\geq5$)~\cite{Kamionkowski:1992mf,Holman:1992us,Barr:1992qq,Ghigna:1992iv} are expected to exist, where $\Phi=(\phi/\sqrt{2})e^{ia/F_{a}}$ the PQ complex scalar and $M_{P}\simeq2.4\times10^{18}{\rm GeV}$ the reduced Planck mass. Without sufficient suppression, the contribution to the axion potential from these operators can easily shift $\bar{\theta}_{\rm min}$ and thus they pose a challenge to resolving the Strong CP puzzle via the QCD axion.

This situation tempts us to gauge the axionic shift symmetry to preclude such problematic perturbative higher dimensional operators. In the dual description, the axion is dualized to a 2-form anti-symmetric field $B_{\mu\nu}$, and the global axionic shift symmetry manifests as a gauge symmetry $B_{\mu\nu}\rightarrow B_{\mu\nu}+\partial_{[\mu}\lambda_{\nu]}$. One can then translate the Strong CP problem as the existence of a non-zero 4-form arising as the field strength of the QCD 3-form that emerges in the infrared limit. This 4-form gets Higgsed via a St{\"u}ckelberg mechanism that couples the 2-form and 3-form~\cite{Dvali:2005an,Dvali:2022fdv}. 

Despite the advantages of this picture, a quality issue can still arise if the QCD axion couples to additional 3-forms beyond the QCD Chern-Simons (CS) 3-form. Nevertheless, when compared to the PQ picture, the QCD axion quality issue is under a better control in the dual description in that underlying physics spoiling $\bar{\theta}_{\rm min}=0$ is unique. This uniqueness leads to a straightforward resolution to the axion quality problem in the dual description: one simply Higgses all the 3-forms coupled to the QCD axion via the inclusion of additional two-forms~\cite{Dvali:2005an,Burgess:2023ifd}. This simple and clear perspective toward the quality issue serves as an appealing point of the dual description in comparison with the PQ picture. The idea of multiple two forms enhancing the quality of the QCD axion in the dual description may motivate the axion-like particles (ALP) even if we restrict ourselves to the low energy effective theories.

This resolution of the QCD axion quality problem has the further attractive feature that it can be motivated from string theory. Multiple 3-forms are not only permitted from an effective field theory viewpoint, they arise naturally and plentifully in string compactifications~\cite{Bousso:2000xa}. Furthermore, Kaluza-Klein reduction of p-forms gives rise to a plethora of axions, giving rise to the so-called Axiverse~\cite{Arvanitaki:2009fg,Cicoli:2012sz}. One can then view the above mechanism as a natural solution to the QCD axion quality problem in the context of the string Axiverse with 3-forms included. 

In this paper, motivated by this observation, we study intriguing implications of the multiple axion scenarios resulting from the high quality of the QCD axion. After reviewing the axion quality problem in the dual picture in Sec.~\ref{sec:axion qualtiy}, we discuss phenemenological implications for axions in general in Sec.~\ref{sec:implication}. Sec.~\ref{subsec:QCD axion} is dedicated to QCD axion phenomenology. This involves a sub-eV and a heavy (TeV scale) ALP coupling to the SM photon which is induced due to the mass mixing with the QCD axion. In Sec.~\ref{subsec:Quintessence axion}, we discuss how the two form dual description can be applied to model-building for a quintessence axion dark energy and the advantage of the framework in explaining the observed cosmic birefringence.

\section{Axion Quality in Dual Description}
\label{sec:axion qualtiy}
\subsection{QCD Axion Quality Problem}
\label{sec:axion qualtiy2}
In the dual formulation of the QCD axion~\cite{Dvali:2005an}, 
the PQ pseudo-scalar is exchanged with a 2-form field $B_{\mu\nu}$. To give a mass to the 2-form, one introduces a 3-form $C$. Since the field strength $G= dC$ is a 4-form, it is non-dynamical in $4d$ spacetime, but its existence implies a uniform electric field.
Furthermore, one can construct the combination $(dB - mC)^2$, with $m$ a dimensionful parameter. This is invariant under $B\rightarrow B+m\Omega$ with $C\rightarrow C+d\Omega$ and is thus a gauge invariant mass term. One can think of this as a higher-form version of the St{\"u}ckelberg mechanism that screens the 4-form electric field.

To apply this picture to solving the Strong CP problem, one identifies the 3-form $C$ with the QCD 3-form that emerges in the IR~\cite{Luscher:1978rn}. When we define the charge density $q(x)\equiv(1/16\pi^{2}){\rm Tr}[F_{\mu\nu}\widetilde{F}^{\mu\nu}]$ from the $SU(3)_c$ field strength, its expectation value with respect to the $\theta$-vacua satisfies $\langle q(x)\rangle\propto\sin\theta$ in the dilute gas approximation~\cite{Callan:1977gz,Callan:1978bm}. The important observation is that the gauge invariant mass term for $C$ enabled by the mixing with the 2-form $B$ removes the pole in the correlator of $C$. This leads to the screening of the 4-form electric field $dC\propto\langle q(x)\rangle$ and thus vanishing $\theta$. This is not the case when the CS 3-form $C_{\mu\nu\lambda}$ is massless in the absence of the 2-form~\cite{Luscher:1978rn}.

However, the axion's capability to maintain CP conservation in the QCD sector can be spoiled if there is an additional 3-form field to which the axion couples aside from the QCD CS 3-form $C_{\mu\nu\lambda}$~\cite{Dvali:2005an}. The extra 3-form $E_{\mu\nu\lambda}$ may be, for instance, a CS 3-form field of a hidden strongly coupled gauge theory or the gravitational CS 3-form whose field strength is $R\widetilde{R}=(\epsilon^{\mu\nu\rho\sigma}/2)R_{\mu\nu\alpha\beta}R_{\rho\sigma}^{\alpha\beta}$, with $R_{\mu\nu\alpha\beta}$ the Riemmann curvature tensor. If the 2-form field transforms as $B_{\mu\nu}\rightarrow B+m'\Omega'$ under the gauge transformation $m'E\rightarrow m'(E+d\Omega')$, $B$ can couple to $E$. When there are fewer 2-forms than 3-forms in the theory, not all of 3-forms can be Higgsed. This prevents the full screening of the QCD 4-form electric field, raising the question of if $\theta_{\rm QCD}$ resulting from the partial screening is small enough. 

To quantify the quality of the QCD axion in this problematic situation, consider the low energy QCD below the scale $\Lambda_{\rm QCD}\simeq0.2{\rm GeV}$ where QCD becomes strongly coupled. The theory is described by~\cite{Burgess:2023ifd}
\bea
\mathcal{L}(C,E,B) &\supset&- \frac{1}{2\cdot 4!} \Bigl( H_{\mu\nu\lambda\rho} H^{\mu\nu\lambda\rho} + K_{\mu\nu\lambda\rho} K^{\mu\nu\lambda\rho} \Bigr)\cr\cr
&& -\frac{1}{2\cdot 3!} G_{\mu\nu\lambda}   G^{\mu\nu\lambda}   - \frac{1}{3!} \, \frac{\epsilon^{\mu\nu\lambda\rho}}{f_{a}}   G_{\mu\nu\lambda}   J_\rho\cr\cr
&&- \frac{1 }{4!}  \epsilon^{\mu\nu\lambda\rho}  \left(\theta_{\rm QCD} \Lambda_{\rm QCD}^2 H_{\mu\nu\lambda\rho}  + \theta_{h} \Lambda_{h}^2   K_{\mu\nu\lambda\rho} \right)\,, \nonumber \\
\label{eq:LCEB}
\eea
where $K =dE$ and $H = dC$ are the gauge invariant field strengths of 
3-forms $E$ and $C$ , respectively, and $G = dB + m_{a} C + M_{a} E$. Here $m_{a}$ and $M_{a}$ are dimensionful parameters. $J_{\rho}$ is the chiral current which carries PQ charge in the PQ picture. For simplicity, here we consider only leading (quadratic) contributions to functional dependence of the action on the field strengths of three forms.

We dualize the theory by enforcing the Bianchi identity $dG=(\Lambda_{\rm QCD}^{2}/f_{a})H+(\Lambda_{h}^{2}/f_{a})K$. The dimensionful parameters $\Lambda_{\rm QCD}$ and $\Lambda_{h}$ guarantee the canonical mass dimension $2$ of $H$ and $K$. Each scale can be interpreted as $\rho^{-1}$ where $\rho$ is a size of instanton of a gauge theory associated with a 3-form, which makes a dominant contribution to the axion potential. Introducing the other dimensionful parameter $f_{a}$ follows from the mass dimension matching in the Bianchi identity. Note that the same dimensionful parameter $f_{a}$ also appears in the coupling of the 2-form field $B_{\mu\nu}$ to the current $J_{\rho}$ in Eq.~(\ref{eq:LCEB}). This is because the above $f_{a}$ in $dG$ will turn out to be the axion decay constant in the PQ picture and axion couples to the chiral fermion current via the derivative coupling with the decay constant $f_{a}$. 

The Bianchi identity can be implemented by adding the following term to the above Lagrangian
\be
-\frac{a}{3!} \, \epsilon^{\mu\nu\lambda\rho} \left(\partial_\mu G_{\nu\lambda\rho} - \frac{1}{4}\frac{\Lambda_{\rm QCD}^{2}}{f_{a}}\,H_{\mu\nu\lambda\rho}- \frac{1}{4}\frac{\Lambda_{h}^{2}}{f_{a}}\,K_{\mu\nu\lambda\rho} \right)\,,
\label{eq:Bianchi}
\ee
where $a$ is the Lagrange multiplier which is to be identified with the QCD axion. As $dG=m_{a}dC+M_{a}dE=m_{a}H+M_{a}K$, we infer the relations $m_{a}=\Lambda_{\rm QCD}^{2}/f_{a}$ and $M_{a}=\Lambda_{h}^{2}/f_{a}$ from the identity.

Integrating out $G$ (equivalently $B$), one obtains the Lagrangian for $a$, $C$ and $E$
\bea
   \mathcal{L}(C,E,a) &=& - \frac{1}{2}\partial_{\mu}a\partial^{\mu}a    - \frac{1}{f_{a}}J^\mu \partial_\mu a-\frac{1}{2f_{a}^{2}}J_{\mu}J^{\mu}\cr\cr
   &&+\,\,(m_{a}a-\theta_{\rm QCD}\Lambda_{\rm QCD}^{2})X+(M_{a}a-\theta_{h}\Lambda_{h}^{2})Y\cr\cr
   &&-\,\,\frac{1}{2}X^{2}-\frac{1}{2}Y^{2}\,,\nonumber\\ 
   \label{eq:Laxion}
\eea
where $X\equiv(1/4!)\epsilon^{\mu\nu\lambda\rho}H_{\mu\nu\lambda\rho}$ and $Y \equiv (1/4!)\epsilon^{\mu\nu\lambda\rho}K_{\mu\nu\lambda\rho}$.

From the equation of motion for $a$ from Eq.~(\ref{eq:Laxion}), we obtain
\be
\frac{\partial V(a)}{\partial a} = m_{a} X + M_{a} Y\,.
\label{eq:dVda}
\ee
On the other hand, the equations of motion of $C$ and $E$ obtained from Eqs.~(\ref{eq:LCEB}) and (\ref{eq:Bianchi}) give us\footnote{In fact, there must be another contribution to the right hand sides in Eq.~(\ref{eq:XY}) which are the integration constants arising from the equations of motion of 3-forms. When a shift transformation rule of these constants is assigned properly, it can cancel a discrete shift of axion, which realizes the shift symmetric property of axion potential~\cite{Dvali:2005an}. In other UV-completed models based on the similar framework~\cite{Kaloper:2008fb,Kaloper:2008qs,Kaloper:2011jz}, the more concrete way to have the desired cancellation is possible by relying on the shift of 3-form conjugate momentum by the amount of a membrane charge. Note that in the dual description, the periodicity of the axion implies a discrete gauge symmetry that protects the mass of the axion~\cite{Kaloper:2008fb,Kaloper:2011jz,Kaloper:2016fbr,DAmico:2017cda}}
\be
X = m_{a}\,a-\theta_{\rm QCD}\Lambda_{\rm QCD}^2,\quad
    Y=M_{a}\,a-\theta_{h}\Lambda_{h}^2 \,.
\label{eq:XY}    
\ee

Now here lies the crux of the axion quality issue in the dual description: Were it not for $Y$, the global minimum $a_{\rm min,X}$ of the axion potential corresponds to $X=0$ via Eq.~(\ref{eq:dVda}) and therefore we infer $a_{\rm min,X}=\theta_{\rm QCD}\Lambda_{\rm QCD}^{2}/m_{a}$ from Eq.~(\ref{eq:XY}). This provides the dynamical axion solution to the strong CP problem. The introduction of the additional 3-form $E$ and its coupling to the axion, however, produces the dependence of $V(a)$ on $Y$ in Eq.~(\ref{eq:dVda}) and thus a new global minimum $a_{\rm min,X,Y}$ of $V(a)$ in the presence of $E$ becomes~\cite{Burgess:2023ifd}
\be
a_{\rm min,X,Y}=\frac{m_{a}\theta_{\rm QCD}\Lambda_{\rm QCD}^{2}+M_{a}\theta_{h}\Lambda_{h}^{2}}{m_{a}^{2}+M_{a}^{2}}\,.
\label{eq:amin}
\ee

The shift in the minimum is 
\bea
\Delta\theta_{\rm min}&=&\frac{a_{\rm min,X,Y}-a_{\rm min,X}}{f_{a}}\cr\cr
&\simeq&
\begin{cases}
\theta_{h}\left(\frac{\Lambda_{h}}{\Lambda_{\rm QCD}}\right)^{4}, & {\rm for}\,\,  m_{a}\gg M_{a}  \,,\\

\theta_{h}-\theta_{\rm QCD},  &{\rm for}\,\,  m_{a}\ll M_{a}  \,,
\end{cases}
\eea
where we have used  $m_{a}=\Lambda_{\rm QCD}^{2}/f_{a}$ and $M_{a}=\Lambda_{h}^{2}/f_{a}$. Hence, unless the shift in the minimum above satisfies
\be
\Delta\theta_{\rm min}\lesssim 10^{-10} \,.
\label{eq:amin2}
\ee
the QCD axion cannot account for the CP invariance of the QCD sector.

Since $\Delta\theta_{\rm min}\simeq\theta_{h}(\Lambda_{h}/\Lambda_{\rm QCD})^{4}$ when $m_{a}\gg M_{a}$ ($\Lambda_{h}\ll\Lambda_{\rm QCD}$), we demand $\Lambda_{h}\lesssim10^{-3}\Lambda_{\rm QCD}$ to satisfy Eq.~(\ref{eq:amin2}). On the contrary, for $m_{a}\ll M_{a}$ ($\Lambda_{h}\gg\Lambda_{\rm QCD}$), constraining $\Lambda_{h}$ does not resolve the spoiled axion quality. Below we restrict ourselves to the later case for discussing the phenomenological implications of the high quality axion in the dual description.\\

%%%%%%%%%%%%%%%%%%%%%%%%%%%%%%%%%%%%%%%%%%%%%%%%%%%%%%%%%%%%%%%%%%%%%%%%%%%%%%%%%%%%%%%%%%%%%%%%%%%%
\subsection{Multiple Axions for QCD Axion Quality} 
We have seen that axion cannot resolve the strong CP problem if there are more 3-forms coupled to the QCD axion than the number of 2-forms in the theory. The straightforward solution to this issue is to simply introduce additional 2-forms into the theory such that all the 3-forms can be Higgsed~\cite{Burgess:2023ifd}.

Suppose we introduce a second 2-form $\tilde{B}$ with $\tilde{G}=d\tilde{B}+M_{b}E$. As with Eq.~(\ref{eq:Bianchi}), we have a Bianchi identity for $\tilde{G}$:
\be
-\frac{b}{3!} \, \epsilon^{\mu\nu\lambda\rho} \left(\partial_\mu \tilde{G}_{\nu\lambda\rho} - \frac{1}{4}\frac{\Lambda_{h}^{2}}{f_{b}}\,K_{\mu\nu\lambda\rho} \right)\,.
\label{eq:Bianchi2}
\ee
After integrating out $G$ and $\tilde{G}$, one obtains a Lagrangian for $a$, $b$, $C$ and $E$. This gives the following relation through the equation of motion of $b$ axion
\be
\frac{\partial V(b)}{\partial b} =M_{b} Y\,.
\label{eq:dVda2}
\ee

Now with two relations for $X$ and $Y$ from  Eqs.~(\ref{eq:dVda}) and (\ref{eq:dVda2}), we can move to a new basis of 3-forms $(\bar{X},\bar{Y})$
\be
\bar{m}_{a}\bar{X} = m_{a} X + M_{a} Y,\quad\bar{m}_{b}\bar{Y}=M_{b} Y\,.
\label{eq:basischange}
\ee
In this basis, the axion potentials satisfy
\be
\frac{\partial V(a)}{\partial a} = \bar{m}_{a}\bar{X}\quad,\quad\frac{\partial V(b)}{\partial b} = \bar{m}_{b}\bar{Y}\,,
\label{eq:dVda3}
\ee
which results in the correspondence between the global minimum $a_{\rm min,\bar{X}}$ ($b_{\rm min,\bar{Y}}$) of $V(a)$ ($V(b)$) and $\bar{X}=0$ ($\bar{Y}=0$). 

From Eq.~(\ref{eq:dVda3}), we can infer that both axions are dynamically driven to $a_{\rm min,\bar{X}}$ and $b_{\rm min,\bar{Y}}$, where $\bar{X},\bar{Y}=0$. Hence, the condition for the QCD axion to solve the strong CP problem, i.e. the correspondence between $\bar{X}=0$ and $a_{\rm min,\bar{X}}$ is restored.

Although our claim was based on the specific case of two 3-forms coupled to QCD axion, this can be readily generalized to the case where multiple 3-forms are coupled to the axion. By a field redefinition of the 3-forms, one can have a QCD axion decoupled from 3-forms other than the QCD 3-form. The key point is that it is possible to change the basis of the 3-forms such that each axion couples to a single 3-form. This will establish the relation 
\be
\frac{\partial V(a_{i})}{\partial a_{i}} = \bar{m}_{a_{i}}\bar{X}_{i},\quad(i=1,2,...)
\label{eq:basischange}
\ee
where $i$ labels an axion $a_{i}$ and a 3-form coupled to $a_{i}$. Therefore, this one-to-one correspondence between each axion potential $V(a_{i})$ and $\bar{X}_{i}$ associates each $a_{\rm min,\bar{X}}$ to $\bar{X}_{i}=0$. When applied to QCD sector, $\bar{X}_{\rm QCD}=0$ guarantees CP conservation.

This shows that the completeness of the axion solution to the strong CP problem requires the matching of the number of 3-forms coupled to QCD axion and that of 2-forms in the theory. Therefore, \emph{the quality issue of QCD axion in the dual description may hint for the presence of extra axions}. The perspective of correlating the extra axions and the quality of QCD axion motivates discussion for potential impacts on QCD axion phenomenology. In Sec.~\ref{sec:implication} we discuss several implications of the framework we presented so far. 

An example of this situation manifested long ago from embedding the QCD axion in the M-theory limit of the E$_8\times$ E$_8$ heterotic string ~\cite{Choi:1985je,Choi:1985bz,Banks:1996ss,Choi:1997an}. Naively, the model-independent axion in these constructions is an excellent candidate for the QCD axion due to its automatic Chern-Simons couplings to gauge fields. However, it couples to both the SM gauge sector as well as the hidden E$_8$ sector (or subgroups thereof, depending on the details of the compactification). This implies that the model-independent axion will generically couple to more than one 3-form. This situation is remedied by including model-dependent axions. One-loop corrections to the gauge kinetic function~\cite{Kaplunovsky:1987rp,Dixon:1990pc} induce a Chern-Simons coupling for the model-depenent axions, and in the large volume limit a linear combination of axions becomes the QCD axion~\cite{Choi:1997an}. The large volume limit has the additional attractive feature of suppressing worldsheet instanton~\cite{Dine:1986zy,Dine:1987bq} effects. 

%%%%%%%%%%%%%%%%%%%%%%%%%%%%%%%%%%%%%%%%%%%%%%%%%%%%%%%%%%%%%%%%%%%%%%%%%%%%%%%%%%%%%%%%%%%%%%%%%%%%

\section{Implications}
\label{sec:implication}

\subsection{QCD Axion}
\label{subsec:QCD axion}
%%%%%%%%%%%%%%%%%%%%%%%%%%%%%%%%%%%%%%%%%%%%%%%%%%%%%%%%%%%%%%%%%%%%%%%%%%%%%%%%%%%%%%%%%%%%%%%%%%%%
\subsubsection{Axion-photon coupling: sub-eV ALP search}
\label{sec:Axion-Photon Coupling}
When the second 2-form $\tilde{B}$ was introduced with $\tilde{G}=d\tilde{B}+M_{b}E$ in the previous section, there arises a mixing between two axions $a$ and $b$ since the set-up couples both axions to the three form $E_{\mu\nu\lambda}$. This mixing will give the $b$ axion an induced coupling to the SM photon provided the theory contains the operator $\propto a(F_{\rm em}\wedge F_{\rm em})$ with $F_{\rm em}$ the field strength of the $U(1)_{\rm em}$ electromagnetic gauge field.

After introducing $b$ axion with $\tilde{G}=d\tilde{B}+M_{b}E$ and integrating out $G$ and $\tilde{G}$, a part of Lagrangian can be written as
\be
\mathcal{L}(X,Y)\supset-\frac{1}{2}X^{2}-\frac{1}{2}Y^{2}+am_{a}X+(aM_{a}+bM_{b})Y\,.
\label{eq:LXY}
\ee
Here $M_{a}=\Lambda_{h}^{2}/f_{a}$ and $M_{b}=\Lambda_{h}^{2}/f_{b}$ with $f_{a},f_{b}$ decay constants in the interaction eigenbasis of axions.

After $X$ and $Y$ ($C$ and $E$) are integrated out, Eq.~(\ref{eq:LXY}) gives\footnote{One may wonder how the axion potential changes in the presence of the kinetic mixing between the two three forms, e.g. $\mathcal{L}(X,Y)\supset\epsilon XY$, in Eq.~(\ref{eq:LXY}) with $\epsilon$ a dimensionless coupling measuring the mixing. But even in this case, $V(a,b)\simeq(1/2)(M_{a}a+M_{b}b)^{2}$ is generated when $\Lambda_{\rm QCD}\ll\Lambda_{h}$ holds.}
\be
V(a,b)=\frac{1}{2}m_{a}^{2}a^{2}+\frac{1}{2}(M_{a}a+M_{b}b)^{2}\,.
\label{eq:Vab}
\ee
In $(X,Y)$ basis of the 3-forms, we see that there is the mixing between $a$ and $b$. 

When $m_{a}\ll M_{a} (\Lambda_{\rm QCD}\ll\Lambda_{h})$ holds, diagonalization of the mass matrix of axions gives
\bea
a&=&-\frac{r}{\sqrt{1+r^{2}}}a'+\frac{1}{\sqrt{1+r^{2}}}b'\,,\cr\cr
b&=&\frac{1}{\sqrt{1+r^{2}}}a'+\frac{r}{\sqrt{1+r^{2}}}b'\,.
\label{eq:eigenbasis}
\eea
where $r=M_{b}/M_{a}$ and the (un)primed basis is the mass (interaction) eigenbasis. And, the mass eigenvalues read
\be
(m_{a'})^{2}\simeq\left(\frac{r}{\sqrt{1+r^{2}}}\right)^{2}m_{a}^{2}\,,\quad(m_{b'})^{2}\simeq (1+r^{-2})M_{b}^{2}\,.
\label{eq:masseigenvalue}
\ee

Substituting Eq.~(\ref{eq:eigenbasis}) into the usual QCD axion-photon coupling we have in PQ picture, i.e. $\mathcal{L} \supset -(g/2)\;a F_{\rm em}\wedge F_{\rm em}$, we find
\be
\mathcal{L}\supset\left(\frac{r}{\sqrt{1+r^{2}}}a'-\frac{1}{\sqrt{1+r^{2}}}b'\right)\frac{g}{8}\epsilon^{\mu\nu\rho\lambda}F_{\mu\nu}F_{\rho\lambda}\,,
\label{eq:axionphoton}
\ee 
where $g$ is a coupling constant with the mass dimension $[g]=-1$. 

Defining 
\be
g_{a\gamma\gamma}\equiv-\frac{gr}{\sqrt{1+r^{2}}},\quad g_{b\gamma\gamma}\equiv\frac{g}{\sqrt{1+r^{2}}}\,,
\label{eq:gagg}
\ee
we see $g_{b\gamma\gamma}/g_{a\gamma\gamma}=r^{-1}$. We identify $g_{a\gamma\gamma}$ with $c_{a}\alpha_{\rm em}/(2\pi F_{a})$ where $c_{a}\simeq-1.92$ and $0.75$ for the KSVZ~\cite{Kim:1979if,Shifman:1979if} and DFSZ~\cite{Dine:1981rt,Zhitnitsky:1980tq} models respectively, $\alpha_{\rm em}=1/137$ is the fine-structure constant and $F_{a}$ is the QCD axion decay constant (in the mass eigenbasis). Similarly, we identify $g=c_{a}\alpha_{\rm em}/(2\pi f_{a})$ and $F_{a}=(\sqrt{1+r^{2}}/r)f_{a}$. Given $r=M_{b}/M_{a}=f_{a}/f_{b}$, this implies that for $r\gg1$, $b$ axion coupling to $U(1)_{\rm em}$ gauge sector is suppressed. The other case $r\ll1$ may be more intriguing in a phenomenological sense.

For $g_{b\gamma\gamma}/g_{a\gamma\gamma}\gg1$ ($r\ll1$), depending on a value of $m_{b'}$ in Eq.~(\ref{eq:masseigenvalue}), the existence of the $b$ axion may theoretically well-motivate axion-like particle (ALP) search based on the axion-photon coupling. In this limit, we have $m_{b'}\simeq M_{a}=\Lambda_{h}^{2}/f_{a}$ from Eq.~(\ref{eq:masseigenvalue}) whereas the QCD axion mass becomes $m_{a'}=\Lambda_{\rm QCD}^{2}/F_{a}=(r/\sqrt{1+r^{2}})m_{a}$.

Interesting is to observe that for a given ($m_{a'},g_{a\gamma\gamma}$), even if $g_{a\gamma\gamma}$ is too small for the current and future experimental sensitivity to reach, there arises a new target parameter space of searching for a sub-eV $b$ axion with $g_{b\gamma\gamma}=g_{a\gamma\gamma}/r$. Given $f_{a}/f_{b}=r\simeq f_{a}/F_{a}$ for $r\ll1$, requiring a sub-Planckian QCD axion decay constant, i.e. $F_{a}\lesssim M_{P}$ gives us $f_{a}/r\lesssim M_{P}$. Then insofar as $\Lambda_{\rm QCD}<\Lambda_{h}\lesssim\sqrt{r}10^{4.5}{\rm GeV}$ holds, $b$ axion mass becomes sub-eV scale, but still larger than that of QCD axion. 

With the $b$ axion being characterized by $m_{b'}/m_{a'}\simeq r^{-1}(\Lambda_{h}/\Lambda_{\rm QCD})^{2}$ and $g_{b\gamma\gamma}=g_{a\gamma\gamma}r^{-1}$, we expect the target parameter space for $b$ axion to be located below the QCD axion band in ($m_{a'},g_{a\gamma\gamma}$). This motivates searches for ALPs in the future even in the event of null results for the QCD axion observation via axion-photon coupling.

The above picture is modified if we allow the $b$ axion to couple directly to electromagnetism via an operator $\mathcal{L}\supset -(g_b/2)\; b F_{\rm em}\wedge F_{\rm em}$. If $g_b\sim g$, then the $a'$ and $b'$ axion photon couplings will have the same asymptotic behaviour in the large and small $r$ regimes. This would imply that axion-photon experiments should expect two axions of nearly identical coupling strength, albeit at different mass values.

One could generalize Eq.~(\ref{eq:LXY}) to even more axions and 3-forms. So long as the additional axions do not couple to the QCD 3-form and the axion $a$ does not couple to the additional 3-forms, then the QCD axion quality problem is still resolved and experiments should expect only two axions. If however the axion $a$ couples to the additional 3-forms, then the last term in Eq.~(\ref{eq:LXY}) must be generalized so that the Lagrangian contains the terms 
\be
    \mathcal{L} \supset \sum_i (a\widetilde{M}_i + b_i \widetilde{N}_i)Y_i
\ee
for some scales $\widetilde{M}_i$ and $\widetilde{N}_i$. Diagonalizing as before, we now have the possibility of many axions coupling to photons, although the strength of the interactions depend on the above scales. Also note that rotating the axions as in Eq.~(\ref{eq:eigenbasis}) will couple additional axions  to QCD. This should manifest as a multi-axion signal in nuclear magenetic resonance (NMR) experiments~\cite{Graham:2013gfa,Budker:2013hfa,Dror:2022xpi}. See also ~\cite{Gavela:2023tzu}  for another framework that anticipates multiple axion signals at NMR and photon-coupling experiments.

\subsubsection{Entropy Production from a heavy ALP}
\label{sec:heavyALP}
A possible scenario is that $f_{a}$ and $f_{b}$ are aligned to be $\mathcal{O}(10^{16}{\rm GeV})$, i.e. $r\simeq1$, from naive expectations of string compactifications~\cite{deCarlos:1993wie,Conlon:2006tq,Svrcek:2006yi,Banks:2003sx,Choi:2006za,Choi:2006qj,Acharya:2010zx}. We find this case interesting since the minimal high quality QCD axion model presented in Sec.~\ref{sec:axion qualtiy} can resolve several cosmological tensions that the string axion scenario suffers from.

The theoretical motivation notwithstanding, as can be seen below~\cite{Fox:2004kb}, such a string axion in the minimal set-up without a tuning for the initial misalignment angle $\theta_{\rm i}\equiv a_{\rm ini}/f_{a}$ gives rise to the axion dark matter relic density exceedingly larger than required
\be
\Omega_{a}h^{2}\simeq(2\times10^{4})\times\theta_{\rm ini}^{2}\times\left(\frac{f_{a}}{10^{16}{\rm GeV}}\right)^{\frac{7}{6}}\,.
\label{eq:Omegaa1}
\ee

Moreover, because of the constraint on the isocurvature perturbation~\cite{Planck:2018vyg} which gives
\be
H_{\rm inf}\lesssim 10^{10}{\rm GeV}\left(\frac{\theta_{\rm ini}}{0.1}\right)\left(\frac{f_{a}}{10^{16}{\rm GeV}}\right)\left(\frac{\Omega_{\rm CDM}h^{2}}{\Omega_{a}h^{2}}\right)\,,
\label{eq:Hinf}
\ee
the overabundance $\Omega_{a}h^{2}\gg\Omega_{\rm CDM}h^{2}$ in turn makes $f_{a}=\mathcal{O}(10^{16}){\rm GeV}$ incompatible with the high scale inflation models unless $\theta_{\rm ini}$ is extremely small. 

This cosmological tensions with a large $f_{a}$ may be reconciled thanks to a heavy $b$ axion enhancing the QCD axion quality. For $\Lambda_{h}\gtrsim10^{10}{\rm GeV}$, one may expect to have $b$ axion as heavy as or heavier than TeV scale. Then, thanks to the $b$ axion coupling to the SM photon in Eq.~(\ref{eq:axionphoton}) with $r\simeq1$, even the perturbative decay rate for $b\rightarrow\gamma+\gamma$ can be large enough to induce the decay of the $b$ axion at a time earlier than the BBN era. Furthermore, such a TeV-scale heavy $b$ axion can be assigned an initial amplitude as large as $f_{b}$. Armed with such a large initial amplitude and TeV scale mass, the $b$ axion is capable of producing a large amount of entropy (via SM photon production) prior to the BBN era.

With these features of an early decay and large energy density, the $b$ axions can thus be exploited to produce entropy (radiation), enabling a sufficient dilution of $\Omega_{a}h^{2}$~\cite{Steinhardt:1983ia,Lazarides:1990xp,Kawasaki:1995vt,Banks:1996ea,Kawasaki:2015pva,Choi:2022btl}.\footnote{A concomitant of the presence of such a heavy scalar is the early matter domination (EMD) era. The entropy release following EDM can be tested by, for instance, the suppression of the stochastic gravitational wave background for the modes entering the horizon before the decay of the heavy scalar~\cite{DEramo:2019tit,Choi:2021lcn}.} Thus, the presence of a heavy $b$ axion can help the string QCD axion be consistent with cosmological constraints without the fine-tuning  of $\theta_{\rm ini}$ or the extension of a QCD axion model.

%%%%%%%%%%%%%%%%%%%%%%%%%%%%%%%%%%%%%%%%%%%%%%%%%%%%%%%%%%%%%%%%%%%%%%%%%%%%%%%%%%%%%%%%%%%%%%%%%%%%
\subsection{Quintessence Axion}
\label{subsec:Quintessence axion}
%\subsubsection{A Quintessence Axion Model and Quality Problem}
\subsubsection{Quintessence Axion Quality Problem and Solution}
The recent observation for the non-zero rotation angle $\beta=\mathcal{O}(0.1)\,$deg of CMB linear polarization~\cite{Minami:2020odp,Eskilt:2022cff} has aroused much interests in an axion-like quintessence~\cite{Choi:2021aze,Gasparotto:2022uqo,Murai:2022zur} (see also \cite{Choi:1999xn,Nomura:2000yk,Ibe:2018ffn,Choi:2019jck,Kim:2022ckq})\footnote{As well as~\cite{Csaki:2001yk,Csaki:2001jk,Csaki:2003ef,Csaki:2004ha} for a different viewpoint on the role of axions in explaining the observed cosmic acceleration.}. However, those quintessence axion models based on the use of an anomalous global $U(1)$ symmetry are challenged by possible corrections to the tiny quintessence mass because the global symmetry the models assume is easily broken by quantum gravity effects~\cite{Hawking:1987mz,Giddings:1988cx,Banks:2010zn}. As a consequence of the quantum gravity lore denying the exact global symmetry, the tiny mass of order $H_{0}$ (the current Hubble expansion rate) for the quintessence as well as a global minimum of the potential for the QCD axion can be easily spoiled by the presence of the global symmetry breaking higher dimensional operators made up of complex scalars~\cite{Kamionkowski:1992mf,Holman:1992us,Barr:1992qq,Ghigna:1992iv}.

Then, in the three-form language, could the tiny mass of a quintessence axion be protected in the same way as the QCD axion? Although the framework for the 3-form gauge theory in the Higgs phase was originally employed in \cite{Dvali:2005an} for studying the QCD axion in the different perspective from the PQ picture, one may apply the same framework to building a model for the quintessence axion dark energy $(q)$ (as an example, see \cite{Kaloper:2008qs}). The necessary modifications include the replacements $\Lambda_{\rm QCD}\leftrightarrow\Lambda_{\rm DE}\simeq2{\rm meV}$, $\theta_{\rm QCD}\leftrightarrow\theta_{\rm DE}$, $f_{a}\leftrightarrow f_{q}\simeq M_{P}$ and $m_{a}\leftrightarrow m_{q}\simeq H_{0}\simeq10^{-33}{\rm eV}$ in Eq.~(\ref{eq:LCEB}). Here, $m_{q}\simeq H_{0}$ is required for satisfying the slow-roll to date, which in turn requires $f_{q}\simeq M_{P}$ for meeting $\Lambda_{\rm DE}^{2}\simeq m_{q}f_{a}$. 

When such a 2-form dual to the quintessence axion mixes with a 3-form other than that for generating the quintessence mass, the quintessence receives an additional contribution to the mass, which may spoil the requisite smallness of $m_{q}$ for the slow-roll today. This implies that the quality issue lurking in the QCD axion is present in quintessence axion models as well. Given that the underlying source causing the problems is identical, we see that if the matching of the numbers of 2-forms and 3-forms are ensured again, the mass generation of the quintessence field can be solely attributed to a coupling to a single 3-form. Namely, the multiple 2-forms (axions) in the 3-form language can protect the quintessence mass in the same manner as the QCD axion. 

As discussed in Sec.~\ref{subsec:QCD axion}, the mixing of quintessence with other ALPs required for protecting the tiny mass may induce a unexpected coupling of the quintessence to other gauge sector. In particular, when the mixing is done with the QCD axion, the quintessence field couples to the SM photon, which might be of a phenomenological interest. Below we discuss how this feature of high quality quintessence axion in the dual formulation can be applied to the phenomenology of the rotation of the CMB linear polarization. 

\subsubsection{A Model Explaining the Cosmic Birefringence}
For the quintessence axion to explain the cosmic birefringence, there must be the coupling between the quintessence field and the SM photon, i.e. $\mathcal{L}\supset-(g_{q\gamma\gamma}/2)\;q F_{\rm em}\wedge F_{
\rm em}$ with $[g_{q\gamma\gamma}]=-1$. Only then does there occur the phase difference in $U(1)_{\rm em}$ gauge fields of different polarization, which causes the rotation of the linear polarization of the CMB~\cite{Harari:1992ea,Fedderke:2019ajk}. 

Given $g_{q\gamma\gamma}=c_{q}\alpha_{\rm em}/(2\pi f_{q})$ with the anomaly factor $c_{q}$ and the decay constant $f_{q}$, the rotation angle so obtained is given by
\be
\beta=0.42\;{\rm deg}\times\frac{c_{q}}{4\pi}\times\frac{q(t_{0})-q(t_{\rm LSS})}{f_{q}}\,,
\label{eq:beta}
\ee
where $\Delta q\equiv q(t_{0})-q(t_{\rm LSS})$ is the field displacement from the last scattering surface (LSS) to today. From the analysis of CMB linear polarization data from Planck and WMAP, the rotation angle $\beta=0.342^{\circ}\substack{+0.094^{\circ} \\ -0.091^{\circ}}$~\cite{Eskilt:2022cff} (3.6$\sigma$) was inferred. 

From the equation of motion of the deviation $\delta q$ of the quintessence field from the hill top, one can obtain $\delta q/f_{q}\propto{\rm exp}(M_{P}^{2}H_{0}t/f_{q}^{2})$. Taking $t\simeq H_{0}^{-1}$ and $f_{q}\simeq M_{P}$, we see that $\Delta q/f_{q}=\mathcal{O}(0.1-1)$. Therefore, if the decay constant $f_{q}\propto g_{q\gamma\gamma}^{-1}$ characterizing the the quintessence coupling to $U(1)_{\rm em}$ sets the period of the quintessence potential, one can account for $\beta=\mathcal{O}(0.1){\rm deg}$ with a reasonable anomaly factor $c_{q}=\mathcal{O}(1-10)$. 

While such a simple and plausible explanation is attractive, there are theoretical criticisms: first is the aforementioned quintessence axion quality issue, and second is a lack of an underlying motivation as to why the quintessence field couples to $U(1)_{\rm em}$. For the later, one may assume a KSVZ-like model with fermions simultaneously carrying charges to both $U(1)_{\rm em}$ and a global $U(1)_{X}$  whose spontaneous breaking gives the quintessence field as its Nambu-Goldstone boson. However, from a UV perspective, there is no reason for introducing such fermions and thus the set-up is rather contrived.\footnote{For the QCD axion, as the quarks contributing to the mixed anomaly $U(1)_{\rm PQ}-[SU(3)_{c}]^{2}$ carry $U(1)_{\rm em}$ charges, the QCD axion coupling to $U(1)_{\rm em}$ is unavoidable. For the quintessence axion, however, if $SU(N)_{X}$ is a non-Abelian gauge theory whose instanton generates the quintessence mass, it is not necessary for a fermion contributing to the mixed anomaly $U(1)_{X}-[SU(N)_{X}]^{2}$ to carry a $U(1)_{\rm em}$ charge. Therefore, unless $SU(N)_{X}$ is identified with $SU(2)_{L}$ in the SM model, assuming fermions carrying charges of both $U(1)_{X}$ and $U(1)_{\rm em}$ is beyond a minimal set-up.}

We propose that the dual formulation of the quintessence axion can address these two questions simultaneously. We have seen that making all the 3-forms massive using multiple (2-form) axions is accompanied by the mixture of axions. This correlation between protecting axion quality and the resulting mixing among axions may naturally cause a quintessence coupling to $U(1)_{\rm em}$ without assuming the multiple contrived model-building ingredients required in the PQ picture. 

For the quality issue, the most worrisome situation would be that both the QCD axion and the quintessence field suffer from couplings to additional 3-forms associated with scales satisfying $\Lambda_{\rm QCD}\ll\Lambda_{h}$. Once there is such a 3-form, say $P_{\mu\nu\lambda}$, the gauge invariant couplings of both axions to $P_{\mu\nu\lambda}$ will modify the minimum of QCD axion potential and the quintessence mass simultaneously. In this general case, making all the 3-forms massive with multiple 2-forms will result in the mixing between the QCD axion and the quintessence field and thereby the coupling between the quintessence field and $U(1)_{\rm em}$ can be induced from the QCD axion-photon coupling $\propto\;-a F_{\rm em}\wedge F_{\rm em}$. 

To see how this works, we study the toy model in which both the QCD axion ($a$) and the quintessence field ($q$) couple to a common extra 3-form $P$ with field strength $S=dP$ and $Z\equiv(1/4!)\epsilon^{\mu\nu\lambda\rho}S_{\mu\nu\lambda\rho}$. We include another axion, $b$, and denote the 2-forms dual to $a$, $q$ and $b$ by $B_{a}$, $B_{q}$ and $B_{b}$, respectively. We also define $C$ as the QCD CS 3-form and $E$ as the 3-form associated with the scale $\Lambda_{\rm DE}$ and attributed to generation of $m_{q}$. Finally, we take $f_{a},f_{q}$ and $f_{b}$ as the decay constants of the axions in the interaction eigenbasis.

Following the construction in~\ref{sec:axion qualtiy2}, we utilize the relevant scales $\Lambda_{\rm QCD}$, $\Lambda_{\rm DE}$, and $\Lambda_{h}$ to construct mass parameters $m_{a}=\Lambda_{\rm QCD}^{2}/f_{a}$, $m_{q}=\Lambda_{\rm DE}^{2}/f_{q}$, $M_{i}=\Lambda_{h}^{2}/f_{i}$ with $i=a,q,b$. Defining composite 3-form $G_{a}=dB_{a}+m_{a}C+M_{a}P$, $G_{q}=dB_{q}+m_{q}E+M_{q}P$ and $G_{b}=dB_{b}+M_{b}P$, we enforce the Bianchi identities $dG_{a}=(\Lambda_{\rm QCD}^{2}/f_{a})H+(\Lambda_{h}^{2}/f_{a})S$, $dG_{q}=(\Lambda_{\rm DE}^{2}/f_{q})K+(\Lambda_{h}^{2}/f_{q})S$ and $dG_{b}=(\Lambda_{h}^{2}/f_{b})S$ with $H=dC$ and $K=dE$.  

One can now generalize Eq.~(\ref{eq:LCEB}) and (\ref{eq:Bianchi}) to include the multiple 2-forms $B_{i}$  and 3-forms $C, E$ and $P$. After integrating out $G_{a}$, $G_{q}$ and $G_{b}$ ($B_{a}$, $B_{q}$ and $B_{b}$), we obtain
\bea
\mathcal{L}(C,E,P,a,q,b)&\supset&-\frac{1}{2}X^{2}-\frac{1}{2}Y^{2}-\frac{1}{2}Z^{2}\cr\cr
&+&am_{a}X+qm_{q}Y\cr\cr
&+&(aM_{a}+qM_{q}+bM_{b})Z\,.
\label{eq:LXYZ}
\eea
where $X\equiv(1/4!)\epsilon^{\mu\nu\lambda\rho}H_{\mu\nu\lambda\rho}$ and $Y \equiv (1/4!)\epsilon^{\mu\nu\lambda\rho}K_{\mu\nu\lambda\rho}$.

Due to the hierarchy $\Lambda_{\rm DE}\ll\Lambda_{\rm QCD}\ll\Lambda_{h}$, the mass parameters satisfy $m_{a},m_{q}\ll M_{a},M_{q},M_{b}$. Then after $C$,$E$ and $P$ are integrated out, Eq.~(\ref{eq:LXYZ}) gives
\be
V(a,q,b)\supset\frac{1}{2}(M_{a}a+M_{q}q+M_{b}b)^{2}\,.
\label{eq:Vaqb}
\ee

From diagonalizing the mass matrix in Eq.~(\ref{eq:Vaqb}), the QCD axion field $a$ in the interaction eigenbasis (unprimed) is found to be the following linear combination of axions in the mass eigenbasis (primed)
\be
a\approx\begin{cases}
\frac{1}{\sqrt{2}}a'-\frac{f_{a}}{2f_{q}}q'+\frac{1}{\sqrt{2}}b'& {\rm for}\,\, f_{q}\gg f_{b}\simeq f_{a} \,,\\
-\frac{f_{a}}{f_{b}}a'-\frac{f_{a}}{f_{q}}q'+b'&{\rm for}\,\,  f_{q}\gg f_{b}\gg f_{a} \,.
\end{cases}
\label{eq:afield1}
\ee

For the case of $f_{q}\gg f_{b}\gtrsim f_{a}$,\footnote{For the other case with  $f_{q}\gg f_{a}\gg f_{b}$, we find that $g_{q\gamma\gamma}\ll c_{a}\alpha_{\rm em}/(2\pi f_{q})$, requiring too a large $c_{a}\gtrsim\mathcal{O}(100)$ for explaining $\beta=\mathcal{O}(0.1){\rm deg}$. Thus, we focus on the case $f_{q}\gg f_{b}\gtrsim f_{a}$.} we see that due to the QCD axion-photon coupling $\mathcal{L} \supset-(g/2)\; aF_{\rm em}\wedge F_{\rm em}$ where $g=c_{a}\alpha_{\rm em}/(2\pi f_{a})$ and $c_{a}$ is the PQ-$U(1)_{\text{em}}$ anomaly coefficient, an induced quintessence-photon coupling arises: $\mathcal{L}\supset -(g_{q\gamma\gamma}/2)\; qF_{\rm em}\wedge F_{\rm em}$ with $g_{q\gamma\gamma}=c_{a}\alpha_{\rm em}/(2\pi f_{q})$. We identify $g_{a\gamma\gamma}=c_{a}\alpha_{\rm em}/(2\sqrt{2}\pi f_{a})$ for the upper case and $g_{a\gamma\gamma}=c_{a}\alpha_{\rm em}/(2\pi f_{b})$ for the upper and lower cases in Eq.~(\ref{eq:afield1}), respectively. 

As a result, the rotation angle of the linear polarization of the CMB induced by the quintessence coupling to $U(1)_{\rm em}$ is given by
\bea
\beta&=&\frac{g_{q\gamma\gamma}}{2}\times[q(t_{0})-q(t_{\rm LSS})]\cr\cr
&\simeq&0.42\; {\rm deg}\times\frac{c_{a}}{4\pi}\times\frac{q(t_{0})-q(t_{\rm LSS})}{f_{q}}\,.
\label{eq:beta2}
\eea
If $c_{a}=\mathcal{O}(10)$ can be realized by multiple fermions contributing to the loop inducing the anomalous coupling of QCD axion to $U(1)_{\rm em}$, $\beta=\mathcal{O}(0.1){\rm deg}$ can be explained for $\Delta q/f_{q}=\mathcal{O}(1)$. 

%%%%%%%%%%%%%%%%%%%%%%%%%%%%%%%%%%%%%%%%%%%%%%%%%%%%%%%%%%%%%%%%%%%%%%%%%%%%%%%%%%%%%%%%%%%%%%%%%%%%

\section{Conclusion}
In the dual formulation of the QCD axion, the potential coupling of the axion to extra 3-forms is the unique fundamental source of the QCD axion quality problem.
Solutions to this problem manifest in one of two ways -- either the coupling to 3-forms is extremely suppressed, or the number of 2-forms and 3-forms in a model must match. While the first scenario is certainly a possibility, we consider the latter scenario less fine-tuned, and so focused on it.

If there is a QCD axion quality problem, the latter scenario may serve as the evidence for the presence of the multiple axion-like particles. We emphasize that correlating the quality problem with multiple axion scenario (axiverse) is one of the distinguishable features of the dual formulation in comparison to the PQ picture.

Importantly, as a consequence of the multiple axion solution, the mixing among axions becomes inevitable. In this work, we have demonstrated how this mixing can provide us with interesting axion phenomenologies.

Given the QCD axion coupling to $U(1)_{\rm em}$, depending on a scale $\Lambda_{h}$ associated with the 3-form spoiling the QCD axion quality, axion-like particles protecting the QCD axion can be either as light as sub-eV scale or as heavy as TeV scale. For sub-eV case in Sec.~\ref{sec:Axion-Photon Coupling}, the QCD axion quality problem may motivate sub-ALP searches exploiting the axion-photon coupling. For the heavy ALP case in Sec.~\ref{sec:heavyALP}, the QCD axion with a decay constant as large as $f_{a}=\mathcal{O}(10^{16}){\rm GeV}$ may be cosmologically safe thanks to the heavy ALP whose decay causes sufficient dilution of the large QCD axion dark matter density. 

We also discussed the usefulness of the dual formulation for models of quintessence axion dark energy. In spite of its simplicity, the explanation of the potentially observed cosmic birefringence by the quintessence axion in the PQ picture may find it difficult to justify the tiny quintessence mass free of corrections due to quantum gravity effects and the coupling between quintessence and $U(1)_{\rm em}$. In contrast, in the dual formulation, these problems can be resolved simultaneously because the mixing among axions resulting from enhancing axion qualities can naturally induce the coupling between quintessence and $U(1)_{\rm em}$. An alternative solution to the observed dark energy in the framework of 4-forms was discussed in~\cite{Bousso:2000xa,Kaloper:2022yiw,Kaloper:2022oqv,Kaloper:2022jpv,Kaloper:2023xfl}. It would be interesting to combine the approach there with the above discussion in solving the QCD axion quality problem, or other phenomenological applications such as those discussed in~\cite{Dvali:2013cpa,Sakhelashvili:2021eid}.

Although we focused on the ALPs coupling to $U(1)_{\rm em}$ in this work, the framework can be also applied to induce QCD axion coupling to dark gauge sectors. For instance, when $b$ axion discussed in Sec.~\ref{subsec:QCD axion} has its own coupling to a dark $U(1)_{D}$ from the outset, this gives QCD axion's coupling to $U(1)_{D}$. This kind of possibility resulting from the QCD axion quality issue in the dual formulation will make the discussion for phenomenological implications richer. We leave exploring other implications for future work.

%%%%%%%%%%%%%%%%%%%%%%%%%%%%%%%%%%%%%%%%%%%%%%%%%%%%%%%%%%%%%%%%%%%%%%%%%%%%%%%%%%%%%%%%%%%%%%%%%%%%

\medskip\noindent\textit{Acknowledgments\,---\,}%
G.C. would like to thank DESY and Imperial College London for their hospitality and support during the completion of this work. J.M.L. would like to thank King's College and the University of Groningen for their hospitality
during the completion of this work. We are grateful to Clifford Burgess, Kiwoon Choi, 
Tony Gherghetta, Thomas Konstandin, Fernando Quevedo, Jérémie Quevillon, Fabrizio Rompineve, Geraldine Servant and Alexander Westphal for discussions. We thank Alexander Westphal and Nemanja Kaloper for comments on a draft of this work.

%%%%%%%%%%%%%%%%%%%%%%%%%%%%%%%%%%%%%%%%%%%%%%%%%%%%%%%%%%%%%%%%%%%%%%%%%%%%%%%%%%%%%%%%%%%%%%%%%%%%

\bibliographystyle{JHEP}
\bibliography{arxiv_1}

%%%%%%%%%%%%%%%%%%%%%%%%%%%%%%%%%%%%%%%%%%%%%%%%%%%%%%%%%%%%%%%%%%%%%%%%%%%%%%%%%%%%%%%%%%%%%%%%%%%%

\end{document}